\documentclass[twocolumn,secnumarabic,showpacs,amssymb, amsmath,tightenlines,aps,prl]{revtex4}
\usepackage{longtable}
\usepackage{bm}
\usepackage{epsfig}
\usepackage{graphicx}
\begin{document}

\title[High-field non-reversible moment reorientation in the antiferromagnet Fe$_{1.1}$Te]{High-field non-reversible moment reorientation in the antiferromagnet Fe$_{1.1}$Te}
\author{W. Knafo$^{1}$, R. Viennois$^{2,3}$, G. Ballon$^{1}$, X. Fabr\`{e}ges$^{1}$, F. Duc$^{1}$, C. Detlefs$^{4}$, J. L\'{e}otin$^{1}$, E. Giannini$^{3}$}

\address{$^{1}$ Laboratoire National des Champs Magn\'{e}tiques Intenses, UPR 3228, CNRS-UJF-UPS-INSA, 143 Avenue de Rangueil,
31400 Toulouse, France.\\
$^{2}$ Institut Charles Gerhardt Montpellier, UMR 5253, Universit\'{e} Montpellier 2 and CNRS, F-34095 Montpellier, France.\\
$^{3}$ D\'{e}partement de Physique de la Mati\`{e}re Condens\'{e}e, Ecole de Physique, Universit\'{e} de Gen\`{e}ve, CH-1211 Gen\`{e}ve, Switzerland.\\
$^{4}$ European Synchrotron Radiation Facility, Bo\^{\i}te Postale 220, F–38043 Grenoble Cedex, France.}

\date{\today}

\begin{abstract}

Magnetization measurements have been performed on single-crystalline Fe$_{1.1}$Te in pulsed magnetic fields $\mathbf{H}\perp\mathbf{c}$ up to 53~T and temperatures from 4.2 to 65~K. At $T=4.2$~K, a non-reversible reorientation of the antiferromagnetic moments is observed at $\mu_0H_R=48$~T as the pulsed field is on the rise. No anomaly is observed at $H_R$ during the fall of the field and, as long as the temperature is unchanged, during both rises and falls of additional field pulses. The transition at $H_R$ is reactivated if the sample is warmed up above the N\'{e}el temperature $T_N\simeq60$~K and cooled down again. The magnetic field-temperature phase diagram of Fe$_{1.1}$Te in $\mathbf{H}\perp\mathbf{c}$ is also investigated. We present the temperature dependence of $H_R$, as well as that of the antiferromagnetic-to-paramagnetic borderline $H_c$ in temperatures above 40~K.

\end{abstract}


% 74.70.Xa 	Pnictides and chalcogenides
% 75.30.Kz 	Magnetic phase boundaries
%(including classical and quantum magnetic transitions, metamagnetism, etc.)
% 75.50.Bb 	Fe and its alloys
% 75.50.Ee 	Antiferromagnetics
% 75.60.-d 	Domain effects, magnetization curves, and hysteresis
% 75.60.Ej 	Magnetization curves, hysteresis, Barkhausen and related effects
% (for hysteresis in ferroelectricity, see 77.80.Dj)
% 75.60.Nt 	Magnetic annealing and temperature-hysteresis effects

\maketitle

Since the discovery of the new class of iron-based superconductors in 2008 \cite{kamihara08}, the superconducting iron chalcogenides Fe$_{1+x}$Te$_{1-y}$Se$_y$ \cite{hsu08} were thoroughly studied because of their simple structure and of the absence of arsenic \cite{ishida09,mizuguchi10,johnston10}. At ambient pressure, the maximal superconducting temperature of 14~K of this system, observed in FeTe$_{0.5}$Se$_{0.5}$ \cite{sales09}, is much smaller than that of other iron-based superconductors. However, it can be raised under pressure to a significantly higher value, i.e., up to 37~K in Fe$_{1.01}$Se \cite{medvedev09}. Similarly to many heavy-fermion superconductors \cite{flouquet06,sarrao07,pfleiderer09,aoki13}, Fe-based superconductors are in the proximity of a quantum paramagnetic-to-antiferromagnetic instability and magnetism is suspected to play a central role for the formation of the superconducting Cooper pairs \cite{tsyrulin12}. Characterizing the magnetic properties of the antiferromagnetic parents of Fe$_{1+x}$Te$_{1-y}$Se$_y$ is thus a key to understand the interplay between magnetism and superconductivity in these systems. In the non-superconducting iron-tellurides Fe$_{1+x}$Te, commensurate bicollinear antiferromagnetism with the wavevector $\mathbf{k}=(\frac{1}{2}~0~\frac{1}{2})$ - made of pairs of chains parallel to $\mathbf{b}$ and ferromagnetically-coupled along $\mathbf{a}$ - develops for $x\leq0.11$ within a first-order transition at $T_N\simeq60-70$~K \cite{johnston10,wen11,bao09,rodriguez11a,rossler11}. This magnetic transition is accompanied by a change of the crystalline structure, which is tetragonal ($P4/nmm$, $\sharp129$) above $T_N$ and monoclinic ($P2_1/m$, $\sharp11$) below $T_N$. While numerous high-magnetic-field studies have been made on Fe-based superconductors to determine their critical field and its temperature dependence (see Ref. [\onlinecite{kida09,braithwaite10,lei10,khim10,klein10}] for Fe$_{1+x}$Te$_{1-y}$Se$_y$), only few have been done so far on their antiferromagnetic parents \cite{tokunaga10,weyeneth11,tokunaga12}. Indeed, in Fe-based superconductors the critical fields $H_{c,2}$ needed to break superconductivity are large, but accessible using non-destructive magnets; in their antiferromagnetic parents, huge magnetic fields are required to break the antiferromagnetic order and drive the sample into a polarized paramagnetic state. In EuFe$_2$As$_2$, the field-induced antiferromagnetic-to-paramagnetic borderline has been followed only at temperatures just below the N\'{e}el temperature $T_N^{Fe}\simeq190$~K (associated to the Fe-ions moments), due to a huge critical field estimated well above 500~T at low-temperature \cite{tokunaga10}. In SmFeAsO, despite the smallness of the N\'{e}el temperature $T_N\simeq5$~K a critical field as high as 40~T is associated to the destabilization of antiferromagnetism at sub-kelvin temperatures [\onlinecite{weyeneth11}]. Recently, the antiferromagnetic-to-paramagnetic borderline of Fe$_{1+x}$Te$_{1-y}$S$_y$ antiferromagnets, for which $45\leq T_N\leq70$~K, has been followed down to temperatures of 20~K in fields up to 60~T \cite{tokunaga12}.

Here we investigate by magnetization measurements the high magnetic field-temperature phase diagram of the iron-telluride Fe$_{1.1}$Te in fields $\mu_0\mathbf{H}\perp\mathbf{c}$ up to 53~T and temperatures from 4.2 to 65~K. A non-reversible reorientation of the antiferromagnetic moments is found to induce a step-like anomaly in the magnetization at $\mu_0H_R=48$~T during the rise of the pulsed field, but not during its fall. We also report the observation of the antiferromagnetic-to-paramagnetic borderline field $H_c$ in temperatures just below $T_N$. From the temperature dependence of $H_R$ and $H_c$, the magnetic field-temperature phase diagram of Fe$_{1.1}$Te with $\mathbf{H}\perp\mathbf{c}$ is extracted.

\begin{figure}[t]
    \centering
    \epsfig{file=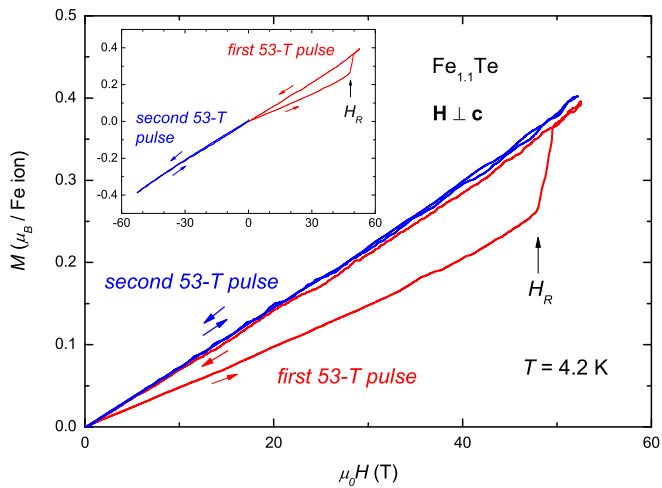,width=87mm}
    \caption{(color online) Magnetization versus magnetic field of Fe$_{1.1}$Te at $T=4.2$~K for $\mathbf{H}\perp\mathbf{c}$, from two successive pulses of magnetic field up to 53~T. An irreversible behavior occurs during the first (zero-field-cooled) pulse, while a reversible behavior is observed during the second pulse (the temperature of 4.2~K was kept constant between the two pulses). The inset shows that a reversible behavior is also observed in a second pulse having an opposite polarity than that of the first pulse.}
    \label{M_4K_2pulses}
\end{figure}

The single crystals of Fe$_{1.1}$Te studied here have been grown by a modified Bridgman method, starting from a nominal cation ration of 1.1:1.0, and all belong to the same batch (cf. Ref. \cite{viennois10} for details on the growth of compounds with slightly different concentrations). Their composition has been determined by chemical analysis using energy-dispersive X-ray spectroscopy, performed on two different samples, and by refinement of X-ray single-crystal diffraction data collected at 180~K from a third sample, using an Oxford X-calibur charge-coupled device diffractometer equipped with a cryojet cooler device from Oxford Instruments and a graphite-monochromatized Mo-$K_\alpha$ radiation source ($\lambda=0.71073$~\AA). This last sample was also characterized by X-ray scattering with an incident energy of 15~keV ($\lambda=0.4$ \AA) on the ID06 beamline at the ESRF (Grenoble, France). The evolution of the $(H00)$ and $(H0H)$ Bragg reflections versus temperature showed a crystalline transition from tetragonal ($P4/nmm$) to monoclinic ($P2_1/m$) symmetry space groups at the temperature $T\simeq60$~K~$\simeq T_N$. Additionally, a splitting of both sets of reflections was observed below T$_N$ indicating the coexistence of two crystallographic domains. Our refined lattice parameters are in good agreement with those reported in the literature \cite{rodriguez11a}:  $a=b=3.809(2)$~\AA\ and $c=6.235(5)$~\AA\ in the tetragonal phase, and $a=3.833(1)$~\AA, $b=3.783(1)$~\AA\, and $c=6.262(5)$~\AA\ ($\beta=89.28~^{\circ}$) in the monoclinic phase. The magnetic susceptibility was measured for magnetic fields $\mu_0\mathbf{H}\perp\mathbf{c}$ of 1~T and $\mu_0\mathbf{H}\parallel\mathbf{c}$ of 0.5~T using a commercial Magnetic Properties Measurement System from Quantum Design. High-field magnetization measurements have been performed for $\mathbf{H}\perp\mathbf{c}$, at temperatures from 4.2 to 65~K, using the compensated-coil technique in a standard 10-mm bore 60-T magnet (30-ms rise, 120-ms fall) at the pulsed-field facility of the Laboratoire National des Champs Magn\'{e}tiques Intenses of Toulouse.

\begin{figure}[b]
    \centering
    \epsfig{file=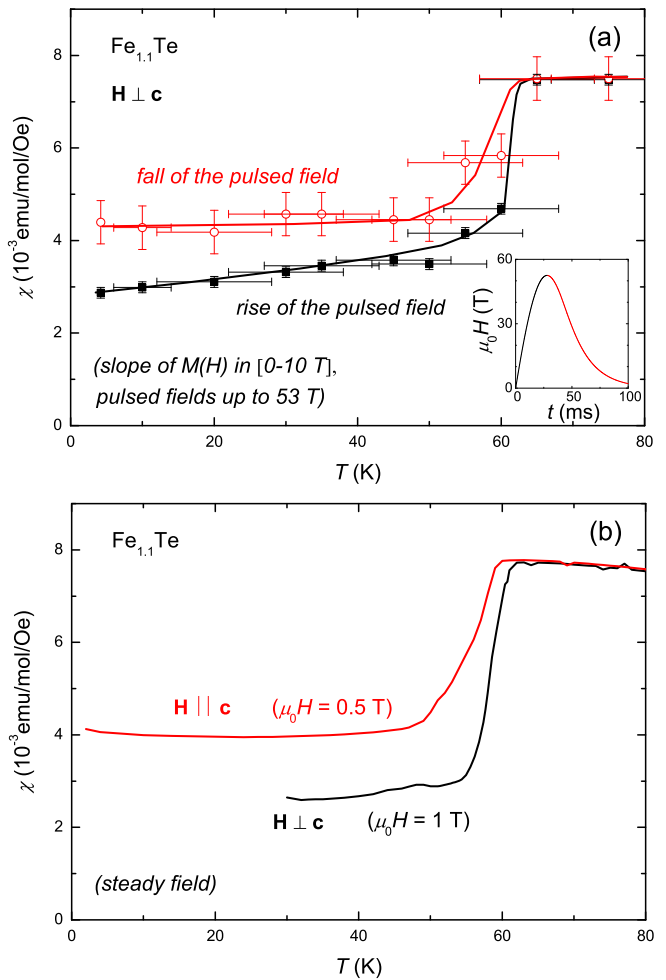,width=87mm}
    \caption{(color online) (a) Magnetic susceptibility versus temperature of Fe$_{1.1}$Te extracted from the slope in the range [0-10~T] of the magnetization versus field $M(H)$ for $\mathbf{H}\perp\mathbf{c}$. The irreversible behavior of the high-field magnetization is emphasized here by the different susceptibilities for the rising and decreasing parts of the magnetic field pulse. The lines are guides to the eyes. (b) Magnetic susceptibility versus temperature of Fe$_{1.1}$Te in steady magnetic fields $\mathbf{H}\perp\mathbf{c}$ of 1~T and  $\mathbf{H}\parallel\mathbf{c}$ of 0.5~T.}
    \label{suscept}
\end{figure}

Fig. \ref{M_4K_2pulses} shows the magnetization versus magnetic field $M(H)$ of Fe$_{1.1}$Te measured at the temperature $T=4.2$~K, i.e., well below the N\'{e}el temperature $T_N\simeq60$~K of the system, in pulsed magnetic fields up to 53~T. In red is shown the magnetization measured during a first pulse of field applied after a zero-field cooling of the sample. A clear non-reversible behavior is observed. During the rise of the pulse, $M(H)$ is linear below 40~T and increases suddenly by a step-like variation $\Delta M\simeq0.1$~$\mu_B$/Fe-ion at $\mu_0H_R\simeq48$~T (defined at the onset of the step), before becoming linear again, with a bigger slope, above 50~T. During the fall of the field, no anomaly is observed in $M(H)$, which is linear with the same slope as observed in rising field above $H_R$. The magnetization measured at the same temperature of 4.2~K during a second pulse of field, again up to 53~T, is shown in blue in Fig. \ref{M_4K_2pulses}. The temperature has been kept constant between the two pulses. During both the rise and the fall of this second pulse, a reversible linear behavior of $M(H)$ is observed and follows exactly the magnetization measured during the fall of the first pulse. As shown in the Inset of Fig. \ref{M_4K_2pulses}, the polarity of the field direction does no matter, since a second pulse of opposite polarity than that of the first pulse also leads to a reversible linear magnetization. At 50~T, the magnetization reaches $\simeq0.4~\mu_B/$~Fe-ion, which is still well below the value of the ordered antiferromagnetic moment $m_{AF}=1~\mu_B/$~Fe-ion reported for Fe$_{1.1}$Te \cite{zaliznyak11} (and much smaller than the antiferromagnetic moment $\simeq2~\mu_B/$~Fe-ion reported for lower Fe-contents \cite{iikubo09,li09,martinelli10}). \begin{figure}[t]
    \centering
    \epsfig{file=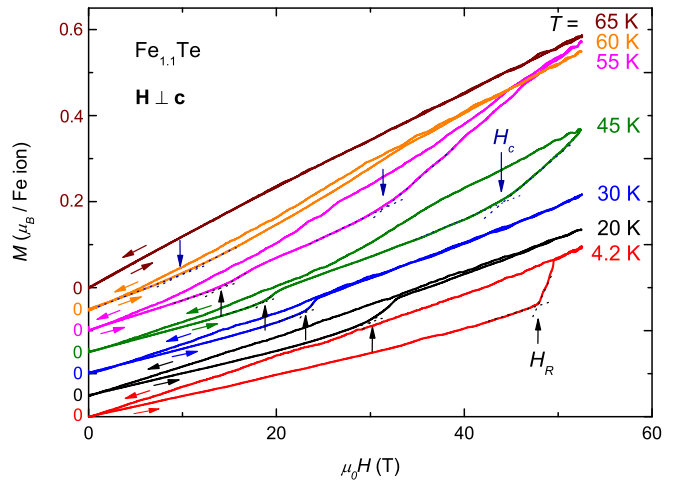,width=87mm}
    \caption{(color online) Magnetization versus magnetic field of Fe$_{1.1}$Te for $\mathbf{H}\perp\mathbf{c}$, at temperatures $T$ from 4.2 to 65~K. All curves were measured during zero-field-cooled pulses of magnetic field, i.e., the temperature was raised above $T_N$ (up to 80~K) and then decreased to its target value before each pulse. For clarity, the origin of each curve has been shifted vertically.}
    \label{M_T_all}
\end{figure}
\begin{figure}[t]
    \centering
    \epsfig{file=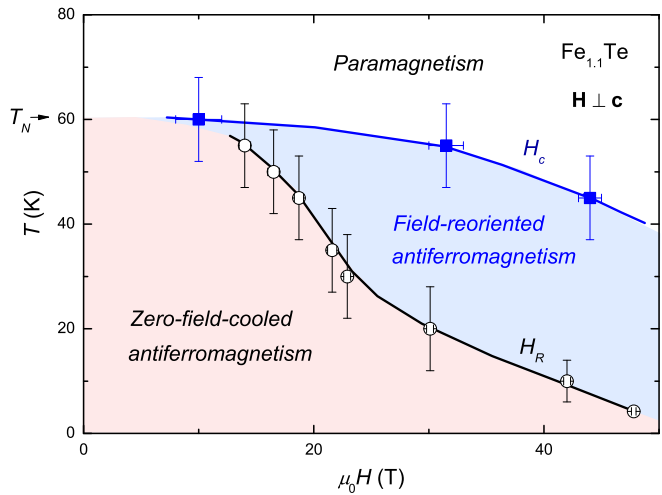,width=87mm}
    \caption{(color online) Magnetic field - temperature phase diagram of Fe$_{1.1}$Te extracted in rising fields $\mathbf{H}\perp\mathbf{c}$ (zero-field-cooling condition). The lines are guides to the eyes.}
    \label{phdiag}
\end{figure} Fig. \ref{M_T_all} shows the magnetization of Fe$_{1.1}$Te at various temperatures $4.2\leq T\leq65$~K. For each of these measurements, the temperature has been systematically raised up to 80~K, i.e., well above $T_N\simeq60$~K, and then decreased before applying the field. All of these zero-field-cooled curves measured for $T<T_N$ show a strong non-reversible behavior. The moment-reorientation field $H_R$ decreases significantly with increasing temperature, and it vanishes at $T=60$~K~$\simeq T_N$. At $T=45$~K, additional strongly hysteretic increase in the slope of $M(H)$ is observed at $\mu_0H_c\simeq44$~T in rising field and $\mu_0H_c\simeq30$~T in falling field (see Fig. \ref{M_T_all}). For $45<T\leq60$~K, $H_c$ decreases rapidly with increasing $T$ and cannot be defined anymore at 65~K. This anomaly marks the destabilization of antiferromagnetism, the moments being polarized paramagnetically above $H_c$. Fig. \ref{phdiag} presents the magnetic field - temperature phase diagram of Fe$_{1.1}$Te extracted in rising fields $\mathbf{H}\perp\mathbf{c}$ under a zero-field-cooled condition, in the extended magnetic field and temperature ranges [0-50~T] and [0-80~K], respectively. Whereas the moment reorientation field $\mu_0H_R$ reaches approximately 50~T at low temperature, the critical antiferromagnetic field $\mu_0H_c$ is only observed down to 45~K. A rough extrapolation of the $H_c$ versus $T$ line indicates that $H_c$ is expected to reach approximately 80-100~T at low-temperature. Field-induced antiferromagnetic-to-paramagnetic borderlines similar to that found here at $H_c$ have already been reported for EuFe$_2$As$_2$ [\onlinecite{tokunaga10}], SmFeAsO [\onlinecite{weyeneth11}], and Fe$_{1+x}$Te$_{1-y}$S$_y$ [\onlinecite{tokunaga12}] antiferromagnets. Similarly to the magnetoresistivity measurements performed on Fe$_{1+x}$Te$_{1-y}$S$_y$ in Ref. [\onlinecite{tokunaga12}], and which were related to magneto-elastic effects, our magnetization measurements show a strong hysteresis at the critical field $H_c$, possibly related to magneto-elastic effects too. The temperature dependence of the magnetic susceptibility extracted from the initial slope, as it fits to experimental data in the range [0-10~T], of our $M(H)$ data for $\mathbf{H}\perp\mathbf{c}$ is shown in Fig. \ref{suscept} (a). In this graph, the black full squares correspond to data taken in the rise of the pulse, i.e., in a zero-field-cooled condition, and the red open circles correspond to data taken in the fall of the pulse. For comparison, Fig. \ref{suscept} (b) shows the zero-field-cooled susceptibility extracted in steady fields $\mu_0\mathbf{H}\perp\mathbf{c}$ of 1~T (black line) and $\mu_0\mathbf{H}\parallel\mathbf{c}$ of 0.5~T (red line). Both susceptibilities extracted in the rise of pulsed fields $\mathbf{H}\perp\mathbf{c}$ and measured in a steady field $\mathbf{H}\perp\mathbf{c}$ correspond to a zero-field-cooled condition and have similar temperature dependences. Interestingly, the susceptibility extracted during the fall of pulsed fields $\mathbf{H}\perp\mathbf{c}$ behaves similarly to that measured in a steady field $\mathbf{H}\parallel\mathbf{c}$.

A challenge is now to describe microscopically the magnetic properties of Fe$_{1.1}$Te under a high field. Different kinds of scenarios can be considered. For example, an irreversible spin-flop-like alignment of the antiferromagnetic moments $m_{AF}$ at $H_R$ during the rise of the pulse could lead to an alignment of the moments perpendicular to the field direction. Within this hypothesis, the condition $\mathbf{H}\perp\mathbf{m_{AF}}$ would be fulfilled for both susceptibilities measured in a steady field $\mathbf{H}\parallel\mathbf{c}$ and during the fall of pulsed fields $\mathbf{H}\perp\mathbf{c}$, which would explain their similar temperature dependences. Alternatively, the step-like anomaly in the magnetization at $H_R$ could be driven by a more subtle change of the magnetic structure, as a modification of the magnetic wavevector (cf. Refs. \cite{mignot91,honma98}) and/or the alignment of the moments from one of the two non-equivalent magnetic Fe-sites (cf. Ref. \cite{fabreges08}). The microscopic model should also describe how a remanent moment reorientation is induced once a magnetic field higher than $H_R$ has been applied and then removed. Knowing that a change of the lattice structure almost coincides with the development of antiferromagnetism at $T_N$ \cite{bao09,li09,martinelli10}, magneto-elastic couplings could possibly play a role in Fe$_{1.1}$Te in a high field. Several mechanisms of coupling between the elastic and magnetic degrees of freedom, for example via an orbital ordering \cite{turner09}, a spin-nematic mechanism \cite{cano11}, or based on symmetry considerations \cite{paul11}, have been proposed in the literature. The question is whether such theories could help understanding the non-reversible moment reorientation observed here. To go further experimentally, magnetostriction under pulsed magnetic field (by strain gage \cite{knafo10}, dilatometry \cite{doerr2008}, or diffraction techniques \cite{duc2010}) could lead to valuable information about the field-induced distortion of the lattice. Also, the high-field magnetic structure of Fe$_{1.1}$Te could be determined by neutron diffraction in pulsed field \cite{yoshii09}.

In conclusion, we have studied the magnetic properties of the antiferromagnet Fe$_{1.1}$Te in pulsed magnetic fields $\mathbf{H}\perp\mathbf{c}$ up to 53~T. In addition to the observation, at temperatures close to $T_N$, of the critical field $H_c$ associated to the antiferromagnetic-to-paramagnetic borderline, we have evidenced a non-reversible moment reorientation at $\mu_0H_R\simeq50$~T at low temperature. The irreversible step-like anomaly in the magnetization at $H_R$ could result from a spin-flop-like reorientation of the antiferromagnetic moments, and/or a change of the magnetic structure (wavevectors, ordered moments etc.). Further efforts from both experimental and theoretical sides are now needed to understand the remanence of the high-field reoriented antiferromagnetic phase. Better knowledge of the magnetic properties of the parent compounds Fe$_{1+x}$Te will surely help understanding the role played by magnetism and magneto-elastic coupling for the appearance of superconductivity in Fe$_{1+x}$Te$_{1-y}$Se$_y$, which coincides with an antiferromagnetic-to-paramagnetic quantum instability coupled with a lattice structure modification.

We acknowledge Christoph Meingast for useful discussions and Laure Vendier for technical help at the X-ray facility of the LCC-Toulouse. This work was supported by Euromagnet II via the EU under Contract No. RII3-CT-2004-506239.

\end{document}